\documentclass[numberedappendix]{emulateapj}
\usepackage{apjfonts}
\usepackage{amsmath}
\usepackage{amsbsy}
\usepackage{amssymb}
\usepackage{epsfig}
\usepackage{natbib}

\newcommand{\kms}{\,{\rm km\,s^{-1}}}
\newcommand{\au}{\,{\rm AU}}

\newcommand{\yr}{\,{\rm yr}}
\newcommand{\myr}{\,{\rm Myr}}

\newcommand{\pc}{\,{\rm pc}}

\newcommand{\msun}{\,M_\odot}

\newcommand{\be}{\begin{equation}}
\newcommand{\ee}{\end{equation}}

\newcommand{\bea}{\begin{eqnarray}}
\newcommand{\eea}{\end{eqnarray}}

\newcommand{\ben}{\begin{enumerate}}
\newcommand{\een}{\end{enumerate}}

\begin{document}

\shorttitle{Origin of HD 188753} 
\shortauthors{PFAHL}

%%%%%%%%%%%%%%%%%%%%%%%%%%%%%%%%%%%%%%%%%%%%%%%%%%%%%%%%%%%%%%%%%%%

\submitted{Accepted by ApJ Letters}

\title{Cluster Origin of Triple Star HD 188753 and its Planet}

\author{Eric Pfahl}

\affil{Kavli Institute for Theoretical Physics, Kohn Hall, University
of California, Santa Barbara, CA 93106; pfahl@kitp.ucsb.edu}

%%%%%%%%%%%%%%%%%%%%%%%%%%%%%%%%%%%%%%%%%%%%%%%%%%%%%%%%%%%%%%%%%%%

\begin{abstract}

The recent discovery by M. Konacki of a ``hot Jupiter'' in the
hierarchical triple star system HD 188753 challenges established
theories of giant-planet formation.  If the orbital geometry of the
triple has not changed since the birth of the planet, then a disk
around the planetary host star would probably have been too compact
and too hot for a Jovian planet to form by the core-accretion model or
gravitational collapse.  This paradox is resolved if the star was
initially either single or had a much more distant companion.  It is
suggested here that a close multi-star dynamical encounter transformed
this initial state into the observed triple, an idea that follows
naturally if HD 188753 formed in a moderately dense stellar
system---perhaps an open cluster---that has since dissolved.  Three
distinct types of encounters are investigated.  The most robust
scenario involves an initially single planetary host star that changes
places with the outlying member of a pre-existing hierarchical triple.

\end{abstract}

%%%%%%%%%%%%%%%%%%%%%%%%%%%%%%%%%%%%%%%%%%%%%%%%%%%%%%%%%%%%%%%%%%%


%%%%%%%%%%%%%%%%%%%%%%%%%%%%%%%%%%%%%%%%%%%%%%%%%%%%%%%%%%%%%%%%%%%

\section{INTRODUCTION}\label{sec:intro}

\citet{konacki05} recently reported the detection of a giant planet in
star system HD 188753.  The planet orbits a Sun-like star every 3.35
days and has roughly the mass of Jupiter. Although this orbit is
extremely tight, it is unremarkable among $\simeq$30 known ``hot
Jupiters'' with periods of 2.5--10 days around solitary
stars\footnote{\url{http://vo.obspm.fr/exoplanetes/encyclo/encycl.html}}.
What distinguishes the new discovery is that the planetary host star
(A) is bound to a companion (B) that is itself a compact binary; HD
188753 is thus a hierarchical triple star.  Star A has a mass of
$M_{\rm A} = 1.06\pm 0.07\msun$, and binary star B has constituent
masses of $M_{\rm B1} = 0.96\pm 0.05\msun$ and $M_{\rm B2} = 0.67\pm
0.05\msun$.  Orbit AB has a semimajor axis of $a_{\rm AB} = 12.3\pm
0.04 \au$ and an eccentricity of $e_{\rm AB} = 0.50\pm 0.05$, and B
has parameters $a_{\rm B} = 0.67\pm 0.01\au$ and $e_{\rm B} = 0.10\pm
0.03$.  While the notion of a planet in a triple star is quite
provocative, the most fascinating aspect of HD 188753 is the close
orbit of AB, which seems to challenge existing models of giant-planet
formation.

Planets form in a disk of gas and dust surrounding a nascent star.
Two theories exist for the genesis of Jovian planets (see Bodenheimer
\& Lin 2002 and references therein).  In the core-accretion paradigm,
a solid core of $\simeq$10 Earth masses forms at disk radii of
$\ga$2--3\,AU, where the temperature is low enough for ices to
condense.  The core then accretes a gaseous envelope.  The alternative
picture involves the gravitational fragmentation of the disk into
gaseous protoplanets at radii of $\simeq$5--10\,AU.  When addressing
hot Jupiters, each theory requires some mechanism to drive the
migration of the planet to small radii \citep[e.g.,][]{lin96}.

Complexities and uncertainties inherent in these two models are
exacerbated when the young star is in a binary.  The perturbing
gravitational field of the binary companion acts to (1) tidally
truncate the disk \citep[e.g.,][]{artymowicz94,pichardo05}, and (2)
stir and heat the disk \citep[e.g.,][]{goodman93, nelson00}.  A disk
truncation radius of $\la$3\,AU may preclude the embryonic stage in
the core-accretion scenario \citep{jang05}.  Even if the truncation
radius is $\simeq$10\,AU, the disk may be too hot to allow either core
formation or unstable gravitational collapse \citep{nelson00}.

The source of the conundrum in HD 188753 is that the orbital
parameters of AB imply that a disk around star A would be truncated at
a radius of only $\simeq$1\,AU.  It seems very unlikely that a giant
planet could have grown in such an environment. However, this dilemma
can be avoided if we abandon the tacit assumption that the system has
maintained its current configuration since the time when the stars and
planet formed.

Suppose that star A and the planet have always been paired, and that
the planet was allowed to form by one of the two models sketched above
and then migrate to small radii.  It follows that, at the time of
planet formation, star A was either single or in a much wider binary
than at present. The simplest way to transform either of these states
into HD 188753 is via a strong dynamical interaction, such as a
scattering encounter between a single and a triple star, or two
binaries.  This could only have occurred in a moderately dense stellar
environment.  Different modes of multiple-star formation and
probabilistic arguments regarding the birthplace of HD 188753 are
discussed in \S~\ref{sec:cluster}.  Three plausible dynamical
histories for HD 188753 are introduced in \S~\ref{sec:dynamics}.  In
\S~\ref{sec:numerical}, numerical scattering experiments are used to
gauge the likelihood of each scenario.  The broader relevance of this
work is given in \S~\ref{sec:conclusions}.

\begin{figure*}
  \centerline{\epsfig{file = 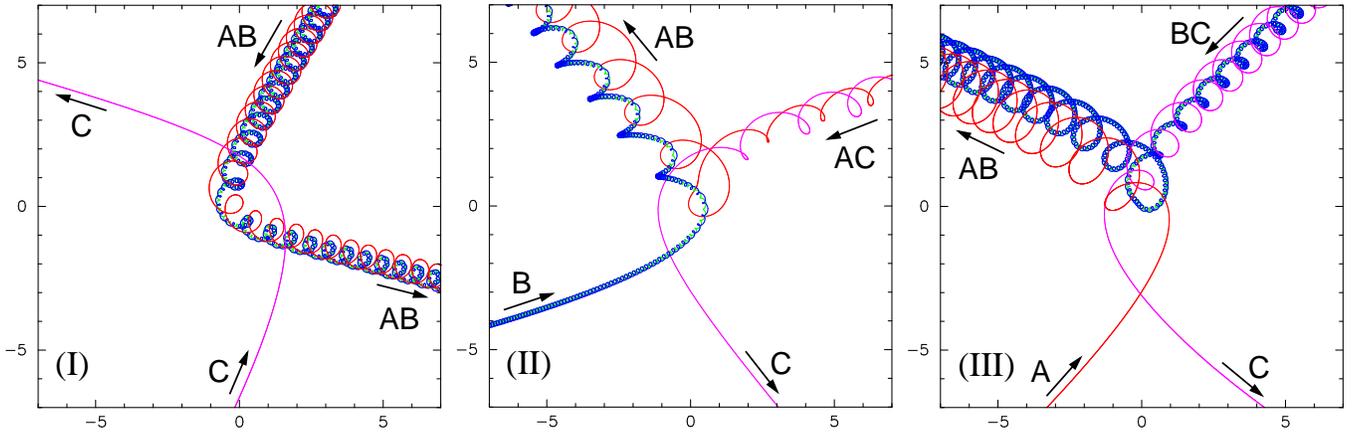, angle = 0, width = 0.99\linewidth}}
  \caption{Numerically generated illustrations of three possible
  dynamical histories for HD 188753. Star A, the planetary host, is
  indicated by the red trajectory, while the tightly curled blue and
  green curves show the components of binary star B.  Star C, shown in
  magenta, is always ejected from the system.
  \label{fig:dynillus}}
\end{figure*}

%%%%%%%%%%%%%%%%%%%%%%%%%%%%%%%%%%%%%%%%%%%%%%%%%%%%%%%%%%%%%%%%%%%

\section{Cluster Properties}\label{sec:cluster}

Stars rarely form in isolation, and so it is quite likely that HD
188753 was born in some flavor of stellar system, either a loose,
transient group of stars or a bona fide open cluster\footnote{The
triple did not form in an old Galactic {\em globular} cluster,
because, among other reasons, star A is too young and massive
\citep[see][]{konacki05}.}.  The salient features of groups and open
clusters are listed below (see Adams \& Myers 2001 for a thorough
discussion).  Once the relevant kinematical parameters are in hand, it
is a simple matter to assess the most probable environmental
conditions for a dynamical origin of HD 188753.

Most stars are born in small groups with $N \sim 10$--$10^2$ members,
stellar densities of $n < 100\pc^{-3}$, velocity dispersions of
$\sigma \la 1\kms$, and evaporation time scales of $T\la 10\myr$.
There is insufficient data on these short-lived groups to develop a
reliable statistical picture of their properties and stellar content.
Roughly 10\% of all stars form in larger systems ($N\sim
10^2$--$10^4$) that would rightfully be called open clusters, for
which there is a wealth of observational data.  Open clusters tend to
have higher densities and somewhat higher velocity dispersions than
groups, and have longer lifetimes.  While some clusters are as old as
$\sim$$10^{10}\yr$, the median age is $\sim$$10^8\yr$.  All but a few
percent of open clusters dissolve in $<$$10^9\yr$, due to the
dynamical effects of internal evolution, the Galactic tidal field, and
close passages by molecular clouds \citep[e.g.,][]{wielen85}.  In open
clusters that have been surveyed for stellar multiplicity, the
estimated binary fraction is $\sim$50\% \citep[e.g.,][]{bica05}.
Although few triples are known in clusters
\citep[e.g.,][]{mermilliod94}, the expected fraction is $\sim$5--10\%,
based on the statistics of field stars near the Sun
\citep[e.g.,][]{duquennoy91}.

A dynamical interaction capable of producing HD 188753 involves at
least one binary (or hierarchical triple) with an orbital separation
of $a \sim 10\au$ (see \S~\ref{sec:dynamics}).  A binary of this size
can be strongly perturbed by a third star if the distance of closest
approach between the star and binary center of mass is $\la$$a$.  For
stellar speeds of $\sigma \sim 1\kms$, we see that $G
M_\odot/\sigma^2\sim 10^3\au \gg a$.  Therefore, the close-encounter
cross section is dominated by gravitational focusing and takes the
characteristic value $\Sigma \sim 2\pi aGM/\sigma^2$, where $M$ is the
total mass of the few-body system.  The probability that the binary
has such an interaction in the lifetime of the stellar system is
\begin{equation}\label{eq:encprob}
P \sim n\Sigma \sigma T 
\simeq 4\times 10^{-3} n_2 a_{10} M_3 \sigma_1^{-1} T_7~,
\end{equation}
where $n_2 = n/100\pc^{-3}$, $a_{10} = a/10\au$, $M_3 = M/3\msun$,
$\sigma_1 = \sigma/1\kms$, and $T_7 = T/10^7\yr$.  Suppose that the
number of similar binaries in any stellar system is a fixed fraction
of the total number of stars.  The net probability for a strong
few-body interaction to occur in a given system would then scale as
$NP$.  A typical open cluster has $N\sim 10^3$ and $T_7\sim 10$
\citep[e.g.,][]{friel95}, making the characteristic value of $NP$ for
a cluster $\sim$100 times larger than for a smaller group.  Even when
we multiply by the additional factor of 0.1, the fraction of stars
formed in open clusters, it is still very probable that HD 188753
originated in a system with $\sim$$10^3$ stars that lived for
$\ga$$10^8\yr$ and has long since evaporated.

%%%%%%%%%%%%%%%%%%%%%%%%%%%%%%%%%%%%%%%%%%%%%%%%%%%%%%%%%%%%%%%%%%%

\section{Dynamical History of HD 188753}\label{sec:dynamics} 

It may seem a daunting task to investigate all possible ways of
forming HD 188753 in a cluster rich in multiple-star systems.
However, three distinct scenarios stand out as being the most likely
(see Fig.~\ref{fig:dynillus} for graphical depictions): (I) hierarchy
AB existed at the time of planet formation, but with a much larger
periastron separation, which was subsequently reduced to its present
value in a strong encounter with a single star; (II) star A was born
with a different, single companion in a wide orbit that was later
exchanged for binary star B; (III) star A was initially single before
being exchanged into a pre-existing hierarchical triple containing B.
Each channel involves a catalyst star (C) that ultimately escapes.  In
all of these interactions, the finite sizes of the planetary orbit,
binary B, and star C are neglected, an idea that is exploited in
\S~\ref{sec:numerical}.

In scenarios I and II, star A is in a binary with B or C at the time
of planet formation.  The semimajor axis $a$ and eccentricity $e$ of
this binary are constrained by the demand that the planet formed by
the core-accretion process or gravitational collapse at a radius of
$\ga$3\,AU in a disk around star A.  \citet{pichardo05} estimate the
disk truncation radius to be $R_t = \lambda a(1-e)^{1.2}$, where
$\lambda$, a function of the stellar masses, decreases from
$\simeq$$1/3$ to $\simeq$$1/5$ as the companion mass increases from
$0.2\msun$ to $2\msun$.  It is not clear how large the disk and binary
orbit must be to allow giant-planet formation to proceed much as it
would around a single star.  The simple constraint adopted in the
numerical study of the next section is that $R_t > 10\au$, which
implies that $a(1-e)^{1.2} \ga 40\au$ in scenarios I and II.

In scenario III, single star A encounters triple BC, where B, a tight
binary with separation $a_{\rm B}$, orbits C with a semimajor axis of
$a_{\rm BC}$ and an eccentricity of $e_{\rm BC}$.  It seems likely
that BC will have long-term stability against disintegration, for
which an approximate condition is \citep{mardling01}
\begin{equation}\label{eq:tripstab}
  a_{\rm BC}(1-e_{\rm BC}) > 2.8 a_{\rm B}
  \left[
	 \frac{M_{\rm BC}}{M_{\rm B}} 
	 \frac{(1+e_{\rm BC})}{\sqrt{1-e_{\rm BC}}}
	 \right]^{2/5}~,
\end{equation}
where $M_{\rm BC} = M_{\rm B} + M_{\rm C}$.  When $a_{\rm B} =
0.67\au$, $M_{\rm B} = 1.63\msun$, and $M_{\rm C} = 1\msun$, the
stability condition is $a_{\rm BC}(1-e_{\rm BC})^{1.2} > 2.3$--$3\au$
for $e_{\rm BC} = 0$--1.  Note that scenario III is much less
constrained than I and II.  It is shown in the next section that
scenario III is the favored channel for producing HD 188753.

%%%%%%%%%%%%%%%%%%%%%%%%%%%%%%%%%%%%%%%%%%%%%%%%%%%%%%%%%%%%%%%%%%%

\section{Numerical Investigation}\label{sec:numerical}

Simulations have been conducted to demonstrate how effective each of
the three scenarios is in producing HD 188753.  Instead of following
all 5 objects---4 stars and 1 planet---the problem has been reduced to
scattering encounters between binaries and single stars.  This
idealization is justified when no pair in the set $\{{\rm A}, {\rm B},
{\rm C}\}$ comes sufficiently close to induce strong tidal
perturbations.  The adopted radii of the objects are the orbital
separation of the planet from A ($0.045\au$), the semimajor axis of B
($0.67\au$), and the radius of a main-sequence star for C,
approximated as $0.005 (M_{\rm C}/M_\odot)\au$.  A scattering
calculation is terminated if the separation of any pair is less than
three times the sum of their radii; the factor of three is a somewhat
arbitrary, but conservative, choice.  In practice, a minor fraction of
the experiments end this way.

Initial conditions for the scattering encounters are generated by
Monte Carlo methods from astrophysically motivated probability
distributions.  The orientations and orbital phases of the incident
binaries are chosen as in \citet{hut83}.  Semimajor axes are drawn
from a logarithmically flat distribution, $p(a) \propto a^{-1}$ ($a =
3$--100\,AU), inspired by binary statistics in the Solar neighborhood
\citep[e.g.,][]{duquennoy91}.  A linear distribution, $p(e) \propto e$
($e = 0$--1), is adopted for the initial eccentricities.  Impact
parameters are distributed uniformly in $b^2$, with a maximum value
large enough to allow all outcomes of interest.  Relative speeds at
infinity, $v$, are picked from a Maxwellian distribution, $p(v)
\propto v^2\exp(-v^2/2\sigma^2)$.  There is no distribution of
catalyst masses, $M_{\rm C}$, that is suitable for all three
scenarios.  As a concrete, but rather arbitrary, choice, $M_{\rm C}$
is drawn from $p(M_{\rm C}) \propto M_{\rm C}^{-1}$ over the range of
0.2--$2\msun$.

\begin{figure}
  \centerline{\epsfig{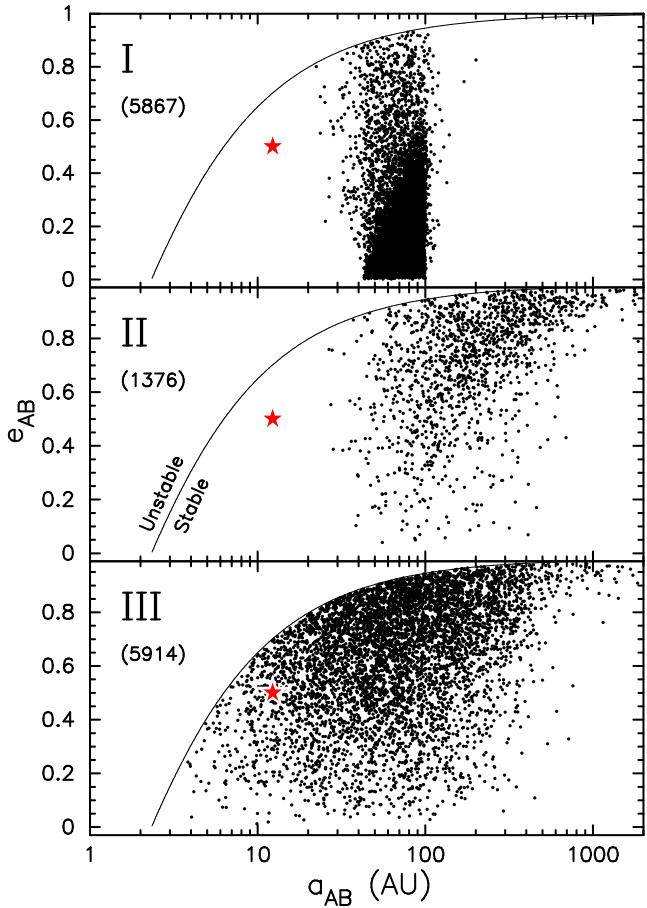}}
  \caption{Each panel shows the semimajor axis and eccentricity of
	 every pair AB that is produced in $10^5$ scattering experiments;
	 the number of dots is given in parentheses.  Red stars mark the
	 orbital parameters of HD 188753.  The curves are the stability
	 boundaries for the triple (see eq.~[\ref{eq:tripstab}]).
	 \label{fig:numerical}}
\end{figure}

Once the parameters of the incident binary have been chosen, they are
checked for consistency with a given scenario.  In scenarios I and II,
$R_t > 10\au$ must be satisfied (by assumption) for the incident
binary containing A.  Scenario III only requires that the incident
triple is stable (eq.~[\ref{eq:tripstab}]).  Each system that passes
this initial test is followed through the scattering interaction.  The
code used to automatically generate and terminate each encounter is
similar in logic to that described in \citet{hut83}.  The equations of
motion are integrated with the time-transformed leapfrog algorithm of
\citet{mikkola99} and \citet{preto99}.

Every time a bound pair AB is left as the final product of an
experiment, the orbital parameters are recorded and the system is
tested for triple stability; unstable systems are cut.
Figure~\ref{fig:numerical} shows the final results of these
simulations.  A total of $10^5$ scattering experiments were computed
for each scenario.  It is immediately apparent in
Fig.~\ref{fig:numerical} that scenarios I and II have considerable
difficulty in producing systems that resemble HD 188753, while the
corresponding efficiency for scenario III is quite high.  If the
minimum disk truncation radius were reduced to, e.g., 5\,AU, scenarios
I and II would be somewhat more profitable.  Nonetheless, scenario III
is clearly the most robust of the three channels, simply because there
are no special hurdles for Jovian-planet formation.

%%%%%%%%%%%%%%%%%%%%%%%%%%%%%%%%%%%%%%%%%%%%%%%%%%%%%%%%%%%%%%%%%%%

\section{Discussion}\label{sec:conclusions}

The obstacles to forming a giant planet in HD 188753 can be overcome
if the triple originated in a moderately dense stellar system and has
a nontrivial dynamical history.  A similar picture might apply to
other close binaries or triples that host a giant planet.  Three
systems are noteworthy in this regard, each with an orbital separation
of $\simeq$20\,AU: $\gamma$ Cephei \citep{hatzes03}; HD 41004 \citep[a
triple containing a brown dwarf;][]{zucker04}; Gliese 86 \citep[has a
white-dwarf secondary;][]{mugrauer05}.  All of these systems could
threaten conventional theories of Jovian-planet formation, but the
compact orbits may have resulted from exchange encounters or other
types of strong dynamical interactions in a stellar group or open
cluster.

Stellar dynamics can play critical roles in the planet-formation
process and in generating variety among single, binary, and triple
stars that host planets.  More than half of all field stars are in
binaries and higher-order multiples, and most of these stars are born
in systems with $\ga$10 members.  Thus, it can be argued that stellar
dynamical considerations are fundamental to a complete understanding
of how planets are created and evolve dynamically, and in deciding how
to search for these worlds.  Earlier studies of planet formation and
dynamics in star clusters
\citep[e.g.,][]{delafuentemarcos97,laughlin98,bonnell01,hurley02}
highlight the complexities and richness of these problems.  However,
much work remains to be done.  Investigations based on scattering
experiments, like those conducted here, should be expanded to include
binary-binary, single-triple, and binary-triple stellar interactions.
Alternatively, detailed $N$-body simulations should be used to study
the fates of planets in systems with $\sim$10--$10^4$ stars, covering
a wide range of initial conditions and assumptions regarding the
statistics of binary and triple stars.

A recent, independent study by \citet{pzm05} uses similar methods of
calculation to those presented here and comes to essentially the same
conclusions regarding the history of HD 188753.

%%%%%%%%%%%%%%%%%%%%%%%%%%%%%%%%%%%%%%%%%%%%%%%%%%%%%%%%%%%%%%%%%%%

\acknowledgements

I thank Phil Arras for numerous valuable discussions and for providing
critical comments on the paper.  I am grateful to the referee, Greg
Laughlin, for asking probing questions about the results of this
study.  This research was supported by NSF grant PHY99-07949.

%%%%%%%%%%%%%%%%%%%%%%%%%%%%%%%%%%%%%%%%%%%%%%%%%%%%%%%%%%%%%%%%%%%

%References

{}

%%%%%%%%%%%%%%%%%%%%%%%%%%%%%%%%%%%%%%%%%%%%%%%%%%%%%%%%%%%%%%%%%%%

\end{document}